 \documentclass[aps,pra,reprint,10pt,twocolumn,nobalancelastpage,flushbottom,
               showpacs,showkeys,
               amsfonts,amssymb,amsmath,longbibliography,lengthcheck]{revtex4-1}
\usepackage[english]{babel}
\usepackage{dcolumn}
\usepackage{bm}
\usepackage{verbatim} % for multiple-line comment
\usepackage[mathscr]{euscript}
\usepackage{graphics,graphicx}
\usepackage{color,xcolor}
\definecolor{dark-blue}{rgb}{0,0,0.875}
\definecolor{dark-green}{rgb}{0,0.625,0}
\definecolor{dark-red}{rgb}{0.875,0,0}
\usepackage[colorlinks=true,linkcolor=dark-red,filecolor=dark-blue,
            citecolor=dark-green,urlcolor=dark-blue,breaklinks=true]{hyperref}
\def\dfrac{\displaystyle\frac}  % Use \displaystyle in the beginning for bigger expressions.
\newcommand{\eps}{\varepsilon}
\begin{document}

\title{Multipolar nonlinear nanophotonics}

\author{\firstname{Daria} \surname{Smirnova}} 
\author{\firstname{Yuri~S.} \surname{Kivshar}} \email[ Corresponding author: ]{ysk@internode.on.net}

\affiliation{Nonlinear Physics Centre, Australian National University, Canberra ACT 2601, Australia}

\date{August 3, 2016}

\begin{abstract}
Nonlinear nanophotonics is a rapidly developing field with many useful applications for a design of nonlinear nanoantennas, light sources, nanolasers, sensors, and ultrafast miniature metadevices. A tight confinement of the local electromagnetic fields in resonant photonic nanostructures can boost nonlinear optical effects, thus offering versatile opportunities for subwavelength control of light. To achieve the desired functionalities, it is essential to gain flexible control over the near- and far-field properties of nanostructures. Thus, both modal and multipolar analyses are widely exploited for engineering nonlinear scattering from resonant nanoscale elements, in particular for enhancing the near-field interaction, tailoring the far-field multipolar interference, and optimization of the radiation directionality. Here, we review the recent advances in this recently emerged research field ranging from metallic structures exhibiting localized plasmonic resonances to hybrid metal-dielectric and all-dielectric nanostructures driven by Mie-type multipolar resonances and optically-induced magnetic response.
\end{abstract}

\maketitle

%------------------------------------------------------------------------------%
% \section*{Main part}\label{sec:intro}
% \textbf{ }
\begin{comment}
\textbf{Nonlinear nanophotonics is a rapidly developing field with many useful applications for a design of nonlinear nanoantennas, light sources, nanolasers, sensors, and ultrafast miniature metadevices. A tight confinement of the local electromagnetic fields in resonant photonic nanostructures can boost nonlinear optical effects, thus offering versatile opportunities for subwavelength control of light. To achieve the desired functionalities, it is essential to gain flexible control over the near- and far-field properties of nanostructures. Thus, both modal and multipolar analyses are widely exploited for engineering nonlinear scattering from resonant nanoscale elements, in particular for enhancing the near-field interaction, tailoring the far-field multipolar interference, and optimization of the radiation directionality. Here, we review the recent advances in this recently emerged research field ranging from metallic structures exhibiting localized plasmonic resonances to hybrid metal-dielectric and all-dielectric nanostructures driven by Mie-type multipolar resonances and optically-induced magnetic response.}
\end{comment}

\section*{Introduction}

Modern nanophotonics aims toward the efficient light manipulation at the nanoscale and the  design of ultrafast compact optical devices for fully-functional photonic circuitry integrable with the state-of-art nanoscale electronics~\cite{NovotnyBook,Shalaev2007_book, Brongersma2007_book,r1c, Menon2010,Miller2010}.
Within decades of fruitful developments, nanophotonics has become a prominent area of research with applications ranging from the high-performance data processing and optical computing to super-imaging and biosensing~\cite{Maier2007_book,Cao2009, Kawata2009,Giessen2011,Maradudin2014_book,sensors}.
Many branches of nanophotonics commonly rely on strong light-matter interaction in the resonant subwavelength structures due to the enhanced near fields~\cite{Schuller2010}, achieved through the excitation of trapped electromagnetic modes, sustained by the nanoelements, or localized geometrical resonances~\cite{Klimov2012_book}.
Resonant tight field confinement is indispensable particularly for the efficient intensity-dependent effects with practically important nonlinear optical applications, such as frequency conversion, wave-mixing, Raman scattering, self-action, and all-optical switching. While in bulk macroscopic media efficiencies of nonlinear optical phenomena are largely determined by the intrinsic nonlinear properties and phase-matching requirements~\cite{Boyd}, efficiencies of nonlinear processes and effective susceptibilities, which may be assigned to nanostructured materials to characterize their nonlinear response, can be significantly increased through enhanced local fields at resonances. Therefore, nonlinear response significantly depends on localized resonant effects in nanostructures, allowing for nonlinear optical components to be scaled down in size and offering exclusive prospects for engineering fast and strong optical nonlinearity and all-optical light control at the nanoscale.

The present-day nanophotonics has greatly advanced in the manufacturing and study of open nanosize optical resonators
that are actually resonant {\em nanoantennas}, which couple localized electromagnetic field and freely propagating radiation~\cite{Novotny2011,Krasnok2013}. Metal nanoparticles~\cite{Klimov2012_book} sustaining %high-$Q$ 
surface plasmon-polariton modes as well as dielectric counterparts with high enough index of refraction~\cite{Staude2013,Krasnok2014_chapter} are utilized as such resonators, and serve for applications in nonlinear diagnostics and microscopy. In turn, two- and three-dimensional clusters and regular arrays of such nanoparticles, employed as unit cells, resonantly responding to IR radiation and visible light, constitute optical metasurfaces and metamaterials (artificial media with pre-designed properties) by analogy with microwave systems~\cite{Sarychev2007_book,Boltasseva2008,Shalaev2010_book}.

{\em Multipole decomposition.} Generally, in the problems of both linear and nonlinear scattering at arbitrary nanoscale objects, multipole decomposition of the scattered electromagnetic fields provides a transparent interpretation for the measurable far-field characteristics, such as radiation efficiency and radiation patterns, since they are essentially determined by the interference of dominating excited multipole modes~\cite{Jackson1965,BohrenBook,Dadap2004,Petschulat2009,Liu2009,Mhlig2011,Gonella2011,Grahn2012}.

In terms of the electric $a_{{E}}$ and magnetic $a_{{M}}$ scattering % $a_{{E}}(l,m)$ and magnetic $a_{{M}}(l,m)$ multipole
coefficients, the total time-averaged scattered power (energy flow)
% radiated energy flux
is given by
\begin{equation}\label{eq:Wsca}
W_s =\dfrac{\pi |E_0|^2}{2\eta k^2} \sum\limits_{l=1}^{\infty} \sum\limits_{m=-l}^{l} {(2l+1) \left( |a_E(l,m)|^2 + |a_M(l,m)|^2 \right)},
\end{equation}
revealing the input of each multipolar excitation (namely, dipole at $l=1$, quadrupole at $l=2$, octupole at $l=3$, etc). Here, $k=\omega\sqrt{\eps\mu}$ % \equiv k_0 \sqrt{\eps_{r}\mu_{r}}
is the wavenumber in the medium, $\eta = \sqrt{\mu/\eps}$ is the impedance, $E_0$ is an electric field amplitude, $l$ and $m$ are angular momentum (orbital) and magnetic  quantum numbers, respectively. Equation~(\ref{eq:Wsca}) is written in SI units under the normalization accepted in Ref.~\cite{Grahn2012}. To describe the radiation from arbitrary localized sources, the multipole coefficients $a_{E,M}$ can be retrieved either through the volume integration of the source current density distribution, or by
taking angular integrals of radial (or angular) components of the numerically pre-calculated electromagnetic
fields with spherical harmonics over a spherical surface enclosing a scatterer.
If no analytical solution can be obtained, the field at the surface of such a sphere can be found with the use of numerical codes or modern commercial full-wave solvers.

Assuming the far-field asymptotic of the outgoing spherical wave,
the field expansion into the vector spherical harmonics ${\bf{X}}_{l,m}( {\theta ,\phi })$ recovers the directional dependence of the radiation, 
% as follows   https://ru.sharelatex.com/learn/Aligning_equations_with_amsmath
\begin{multline}\label{eq:RadPat}
\dfrac{d P (\theta, \varphi)}{d \Omega} =\dfrac{\pi |E_0|^2}{2\eta k^2}
\Biggl\lvert \sum\limits_{l=1}^{\infty} \sum\limits_{m=-l}^{l}  (-i)^{l+1} (2l+1) \\
\cdot \left( i^{l} a_M(l,m) {\bf{X}}_{l,m} + i^{l+1} a_E(l,m)  {\bf {\hat{r}}} \times {\bf{X}}_{l,m}  \right) \Biggr\rvert ^{2},
\end{multline}
defined as a power per unit solid angle $\Omega$ in spherical coordinates $\{ \theta, \varphi \}$, $  \bf {\hat{r}} $ denotes the unit radius vector.
%
% The scattered electric far-field expanded up to $l=2$ % in the far-field % , corresponding to different multipolar orders up to $l=2$,
% is % written as given by~\cite{Jackson1965}
% The expansion of the scattered electric far-field can be expressed by the sum in Cartesian multipoles
% The scattered electric far-field % expanded in Cartesian multipoles can be expressed by the sum
The scattered electric far-field can also be conveniently expressed through the % expansion
decomposition in the Cartesian multipolar terms~\cite{Jackson1965} as a sum
%\begin{equation}
\begin{multline} \label{eq:Hamiltonian}
{\bf{E}}_{\text{sca}} = {\bf{E}}_{\text{ED}} + {\bf{E}}_{\text{MD}} + {\bf{E}}_{\text{EQ}} + \ldots  = \dfrac{k^2}{4 \pi \varepsilon_0}  \dfrac{e^{ikr}} {r}  % \dfrac{e^{i(kr-\omega t)}} {r}
 \Biggl\{ [[ {\bf {\hat{r}}} \times {\bf{p}} ] \times{\bf {\hat{r}}}] \\ - \dfrac{1}{c}[ {\bf {\hat{r}}} \times {\bf{m}} ] -
 \dfrac{ik}{6}[[ {\bf {\hat{r}}} \times {\bf{Q}} ({\bf {\hat{r}}}) ] \times{\bf {\hat{r}}}]  + \ldots \Biggr\} ,
\end{multline}
% \end{equation}
where ${\bf{p}}$, ${\bf{m}}$ and ${\bf{Q}}$ are the electric dipole (ED), magnetic dipole (MD) and electric quadrupole (EQ) leading moments, respectively, and $\eps_0$ is the vacuum permittivity.
% $  \bf {\hat{r}} $ denotes the unit radius vector.
%
Tuning the contributions of different-order multipole moments is used to engineer the scattering and tailor the emission directionality
of optical nanoantennas~\cite{Liu2012,Rodrigo2013,Staude2013,Fu2013,Poutrina2013,Krasnok2014,Smirnova2014,DavidsonII2015, Dregely2014, Liberal2015}. In particular, the so-called first Kerker condition for overlapped and balanced orthogonal electric and magnetic dipoles represents an
example of {\em unidirectional scattering}~\cite{Kerker1983} from a single-element antennas. %zero backscattering
The scattering is interferencially suppressed in either backward or forward ($z$) direction,
if the relation $1/\eps_0 p_x = \pm \eta m_y $ % where $\eps_0$ is the vacuum permittivity,
holds  % the written
for the only non-negligible Cartesian electric $p_x$ and magnetic $ m_y $  dipolar moments induced in a particle along the $x$ and $y$ axes, respectively.
Developing this concept, the directionality of the
scattering can be improved through the interference of properly excited higher-order electric and magnetic modes~\cite{WLiu2014, Hancu2014, Naraghi2015, Alaee2015}.

\begin{figure*}[t!]
% \centerline{\mbox{\resizebox{8.4cm}{!}{\includegraphics{Fig2_Plasmonics_v1}}}}
\centering\includegraphics[width=0.97\linewidth]{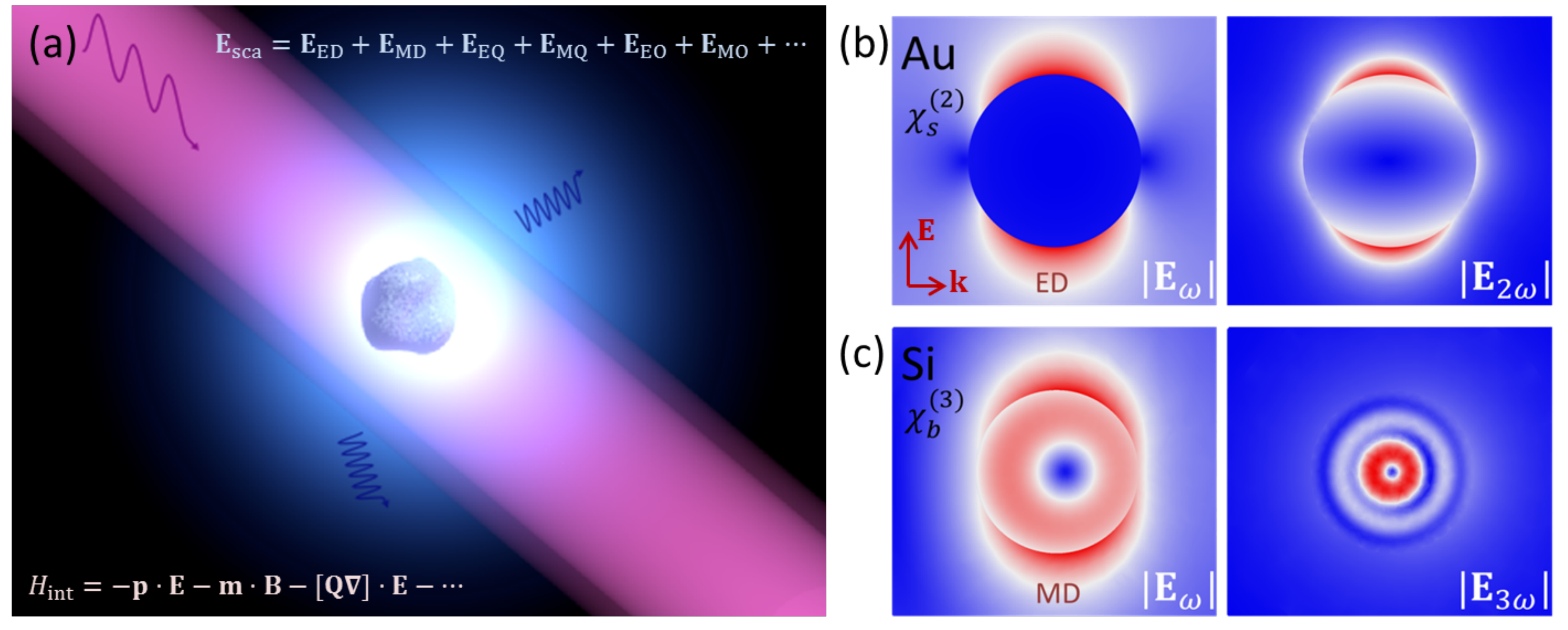} %Fig0_Au_2.png
\caption{
(a) Schematic of nonlinear light scattering by an arbitrarily shaped nanoparticle.
The formulae delineate % (sketch, depicture)
the light-matter interaction Hamiltonian (bottom) and the % (scattered)
nonlinearly generated field (top) expanded in terms of multipole moments.
Illustrations of (b) SHG and (c) THG from a single spherical nanoparticle made of gold and silicon, respectively.
Shown are near-field distributions of the electric field magnitude at the fundamental and harmonic
frequencies. Nonlinear response of a $150$ (600) nm diameter gold (silicon) nanoparticle is driven by ED (MD) mode at wavelength $\lambda_0=0.794$ (2.15) $\mu$m and dominated by the surface (bulk) nonlinear polarization source, characterized by the nonlinear susceptibility tensor $\mathord{\buildrel{\lower3pt\hbox{$\scriptscriptstyle\leftrightarrow$}} \over \chi }^{(2)}_s$ ($\mathord{\buildrel{\lower3pt\hbox{$\scriptscriptstyle\leftrightarrow$}} \over \chi }^{(3)}_b$) of the second (third) order.
Polarization of the exciting % illuminating 
plane wave is specified by red arrows in caption (b).
}
\label{fig:fig1}
\end{figure*}

However, % as discussed e.g. in Refs.~\cite{Roke2004,OBrien2015},
the local-field features and far-field properties are intimately linked in the nonlinear response of nanostructures~\cite{Roke2004,OBrien2015}. A formal way to quantify this effect is based on the Lorentz reciprocity theorem. In particular, this method allows one to predict the metamaterial nonlinearity by using linear calculations~\cite{OBrien2015}, and to express the excitation coefficients of spherical multipoles, the radiated field is decomposed into, through the overlap integrals of the nonlinear source and the respective spherical modes~\cite{thg_dasha}.
In this regard, two key ingredients to realize strong nonlinear response from nanostructures are usually emphasized: the local field enhancement and modal overlaps.
For instance, the efficiency of harmonic generation %
can be strongly enhanced in nanostructures, provided the pump or generated frequency matches the supported resonance~\cite{Boyd}, especially, if the geometry is doubly resonant~\cite{Thyagarajan2012,NavarroCia2012,Aouani2012,Ginzburg2012,Celebrano2015}, %Hayat2007,Banaee2011,
i.e. it sustains resonances at both the fundamental and harmonic frequencies, and spatial distribution of the nonlinear source is such that it strongly couples the corresponding modes.

For metamaterials, the applicability of simple estimates based on the nonlinear oscillator model~\cite{Bassani1998,Poutrina2011} and Miller's law~\cite{Miller1964,Garrett1966} is
limited to the specific cases only~\cite{OBrien2015,Butet2015josab}, e.g. odd in frequency third-harmonic generation~\cite{Hentschel2012, Metzger2012,Metzger2014}.
In fact, the character and efficiency of the nonlinear response in nanostructures are highly affected by many factors~\cite{Czaplicki2015}, including variations in
their size, shape, material filling fractions, quality of samples, interparticle interactions,
spatial symmetries of both geometry and excitation beams~\cite{Bautista2016}, linear scattering properties. Nonetheless, to a large extent, the key governing features may be approached with the analysis
of the unique resonant % linear
behaviour of these artificial materials~\cite{deCeglia2015}, as well as strengths and mutual interference of the leading excited or generated field multipoles~\cite{Bernasconi2016}.

{\em Nonlinear response.} Within the macroscopic description utilizing Maxwell's equations, nonlinear optics employs the nonlinear constitutive relations. For example, in the electric dipole approximation of light-matter interaction, the material response in a nonmagnetic medium can be specified by the nonlinear relationship between the applied electric field ${\bf{E}}$ and induced polarization %(current)
${\bf{P}}$ as
\begin{equation}\label{eq:ConstRel}
{\bf{P}} = \eps_0 \biggl[
  \mathord{\buildrel{\lower3pt\hbox{$\scriptscriptstyle\leftrightarrow$}} \over \chi } ^{(1)}{\cdot} {\bf{E}}  +
  \mathord{\buildrel{\lower3pt\hbox{$\scriptscriptstyle\leftrightarrow$}} \over \chi } ^{(2)}{\rm{:}} {\bf{E}}{\bf{E}}  +  \mathord{\buildrel{\lower3pt\hbox{$\scriptscriptstyle\leftrightarrow$}} \over \chi } ^{(3)} {\scriptscriptstyle{\vdots}} {\bf{E}}{\bf{E}}{\bf{E}} + \ldots \biggl] \text{ , }
\end{equation}
written in the form of an asymptotic expansion in a Taylor series. Here, the first term corresponds to the linear regime at weak excitation fields, and  $\mathord{\buildrel{\lower3pt\hbox{$\scriptscriptstyle\leftrightarrow$}} \over \chi } ^{(N)}$ are the $N$-th order susceptibility tensors of rank $N+1$,
which capture both the polarization dependent nature of the parametric interaction as well as the symmetries of the specific material. Since the optical nonlinearities of natural materials are rather weak, the nonlinear scattering manifests itself at sufficiently strong applied electromagnetic fields, achievable with powerful coherent light sources. %, lasers. 
Given the fact that considerable amounts of electromagnetic energy can be confined to tiny volumes in nanoparticles or even smaller hot spots, they enable downscaling the required optical powers,
because the intensity of SHG/THG processes scales with the fourth/sixth power of the fundamental field strength. 

Symmetry considerations appear to be of particular importance in nonlinear optics. 
For instance, the second-order nonlinear effects are inhibited
in the bulk of such uniform centrosymmetric media, such as plasmonic metals and group IV semiconductors, 
within the electric dipole approximation of the light-matter interaction
because of the symmetry constraints, while no such restriction exists for third-order processes~\cite{Lippitz2005}.
However, the inversion symmetry is broken at interfaces, thus enabling the second-order nonlinear processes from surfaces of isotropic crystals due to the electric-dipole surface contribution to the nonlinear polarization.
Its sensitivity to surface properties is used in probing techniques.
The bulk nonlinear polarization arises from higher-order
nonlocal magnetic-dipole and electric-quadrupole
interactions with light at the microscopic level.
To account for the multipolar orders, the effective light-matter interaction Hamiltonian~\cite{NovotnyBook,Kasperczyk2015} is expanded as such
% in the effective Hamiltonian~\cite{Kasperczyk2015}...%
\begin{equation}\label{eq:Hamiltonian}
H_{\text{int}}  = - {\bf{p}} \cdot {\bf{E}} - {\bf{m}} \cdot {\bf{B}} - [{\bf{Q}} {\bf{\nabla}}]  \cdot  {\bf{E}} - \ldots ,  % + {\bf{Q}} {\rm{:}} \nabla {\bf{E}}  % Akcipetrov
\end{equation}
where the electric dipole ${\bf{p}}$, magnetic dipole ${\bf{m}}$ and electric quadrupole moments ${\bf{Q}}$ are here interpreted as % quantum-mechanical
operators.
% and ${\bf{m}}$ and ${\bf{Q}}$ are of the same order.

\begin{comment}
The aim of this review paper is twofold. First, we demonstrate the importance of the multipolar effects in nonlinear optical processes at the nanoscale when the generation of the specific structured light by the second- or third-harmonic generation processes can change dramatically the structure of the outcome radiation patterns due to the interference of the fields generated by different resonant modes supported by the subwavelength structures.  Second, we emphasize the importance of the magnetic dipole resonances and optically-induced magnetic response in hybrid and all-dielectric nanoscale structures which may lead to a substantial enhancement of nonlinear effects and novel ways to control both directionality and structure of the generated harmonic fields.  Importantly, the physics of resonances in metallic and dielectric nanostructures is quite different, as illustrated in Fig.~\ref{fig:fig1}, and therefore we discuss separately nonlinear effects in plasmonic (metallic), metal-dielectric, and all-dielectric structures where the nonlinear 
response can be dominated by optically-induced magnetic resonances.
% due to the interference of the fields generated by different resonant modes supported by the subwavelength structures. 
% where the response can be dominated by optically-induced magnetic resonances. 
% In this review paper, within the connecting frame of multipolar scattering, we discuss the nonlinear optical effects at the nanoscale. 
\end{comment}

The aim of this review paper is twofold. First, within the connecting frame of multipolar scattering, we discuss the nonlinear optical effects at the nanoscale. Importantly, the physics of supported resonances and material properties in plasmonic and dielectric nanoparticles are different, as illustrated in Fig.~\ref{fig:fig1}, and therefore we consider metallic, metal-dielectric and all-dielectric structures separately.
% electric dipole in deeply subwavelength metal nanoparticles.
Second, we emphasize the importance of the optically-induced magnetic resonances in hybrid and high-index dielectric nanostructures which may lead to a substantial enhancement of the nonlinear response and novel ways to control directionality of the nonlinear radiation.  

\section*{Plasmonic nanoparticles}

To date, the possibilities of nanoscale confinement of light % and operations with photonic flows 
are primarily associated with plasmon-polaritons, which are collective excitations
originating from coupling of the electromagnetic fields to electron oscillations in a metal plasma~\cite{Maier2007_book}. Physics of light interaction with metal structures that are much smaller than the free space wavelength of light constitutes one of the most significant branches of contemporary nanophotonics – {\em nanoplasmonics}. Combining
strong localized surface plasmon resonances and high intrinsic nonlinearities, plasmonic structures offer a unique playground to study a rich diversity of nonlinear phenomena,
including second-harmonic generation (SHG)~\cite{Cao2007,Cao2009,Roke2012,Kauranen2012,Butet2015,Segovia2015}, third-harmonic generation (THG)~\cite{Lippitz2005,Metzger2012,Hentschel2012,Metzger2014}, four-wave mixing (FWM)~\cite{Zhang2013}, multiphoton luminiscence~\cite{Biagioni2012}, self-action effects~\cite{Davoyan2009,Noskov_prl,noskov_njp,Neira2015,Huang2016}.
In a broader scope, the intense electric near-fields may also boost (enhance) %(amplify)
the nonlinear response in the materials brought into the proximity of plasmonic nanoantennas~\cite{Fan2006,alu2,Zhu2013_ITO,Metzger2014ITO,alu2015,Smirnova2015}. Some important aspects of {\em nonlinear plasmonics} were reflected in several recent review papers~\cite{Kauranen2012,Roke2012,30,Butet2015}.

Plasmons are commonly associated with conventional metals (gold, silver, copper and aluminum) possessing a large number of quasi-free conduction electrons which oscillate collectively in response to the applied harmonic field. {\em Electric dipolar resonance} caused by such plasma oscillations in finite structures is most widely exploited for small metallic particles and their composites in nonlinear plasmonics~\cite{Roke2012,Metzger2012,Hentschel2012,Metzger2014}. Noticeably, for deep subwavelength scatterers, even when the excitation wavelength is strongly detuned from the resonance frequency, the electric dipole order dominates over the higher-order excitations in the multipolar decomposition. 

The bulk of plasmonic metals
formed by atoms organized in face-centered cubic lattices exhibits a center of symmetry.
Assuming a particle made of a centrosymmetric isotropic homogeneous medium, the SH polarization source can be written as a sum of
nonlocal-bulk and local-surface contributions,
\begin{equation}\label{eq:P_nl}
{\bf{P}}^{\left( {2\omega } \right)}  = {\bf{P}}_{\text{surf}}^{\left( {2\omega } \right)} +  {\bf{P}}_{\text{bulk}}^{\left( {2\omega } \right)} \text{ , }
\end{equation}
\begin{equation}\label{eq:Psurf1}
{\bf{P}}_{\text{surf}}^{\left( {2 \omega } \right)} = 
% {\varepsilon _0} \;
\varepsilon_0 \mathord{\buildrel{\lower3pt\hbox{$\scriptscriptstyle\leftrightarrow$}} \over \chi } _s^{\left( {2} \right)}{\rm{:}}
{{\bf{E}}^{\left( \omega  \right)}}{{\bf{E}}^{\left( \omega  \right)}} \delta({\bf{r}}-{\bf{r}}_s) \equiv {\bf{P}}_s^{\left( {2\omega } \right)} % (\theta, \varphi)
\delta({\bf{r}}-{\bf{r}}_s) \:, % \right| _{\Sigma_i} % \text{ on } \Sigma \text{ , }
\end{equation} % Eq. (\ref{eq:Psurf1})
\begin{equation}\label{eq:Pb}
{\bf{P}}_{\text{bulk}}^{\left( {2\omega } \right)} = 
\varepsilon_0 \mathord{\buildrel{\lower3pt\hbox{$\scriptscriptstyle\leftrightarrow$}} \over \chi } _b^{\left( {2 } \right)}{\rm{:}}
{{\bf{E}}^{\left( \omega  \right)}} \nabla {{\bf{E}}^{\left( \omega  \right)}} \:,
\end{equation}
where ${\bf{r}}_s$ defines the surface,  ${\mathord{\buildrel{\lower3pt\hbox{$\scriptscriptstyle\leftrightarrow$}} \over \chi } _s^{\left( {2 } \right)}}$ and ${\mathord{\buildrel{\lower3pt\hbox{$\scriptscriptstyle\leftrightarrow$}} \over \chi } _b^{\left( {2} \right)}}$  are the second-order susceptibility tensors.
If the particle surface possesses isotropic symmetry with a mirror plane perpendicular to it, the surface second-order susceptibility tensor ${\mathord{\buildrel{\lower3pt\hbox{$\scriptscriptstyle\leftrightarrow$}} \over \chi } _s^{\left( {2 } \right)}}$ has only three non-vanishing and independent elements, ${{\chi^{\left( {2} \right)}_{ \bot  \bot  \bot }}}$, ${{\chi^{\left( {2} \right)}_{ \bot \parallel \parallel }}}$ and ${{\chi^{\left( {2} \right)}_{\parallel  \bot \parallel }} = {\chi^{\left( {2} \right)}_{\parallel \parallel  \bot }}}$, where the symbols $\bot$ (or $\parallel$) refer to the directions normal (tangential) to the surface~\cite{Heinz} and, therefore, the surface polarization is recast to % is expressed as
\begin{equation}
% \begin{aligned}
{\bf{P}}_{\text{s}}^{\left( {2\omega } \right)} = {\bf{\hat{n}}} ( \chi^{\left( {2} \right)}_{ \bot  \bot  \bot } E'_{n} E'_{n} +    \chi^{\left( {2} \right)}_{\bot \parallel \parallel}  {\bf{E}}'_{\tau} {\bf{E}}'_{\tau}) + 2  \chi^{\left( {2} \right)}_{\parallel  \bot \parallel } E'_{n}{\bf{E}}'_{\tau} \text{ , }
% \end{aligned}
\end{equation}
where ${\bf{E}}'^{(\omega)}$ is the driving field near the interface, $n$ and $\tau$ refer to the normal and tangent components, respectively.
% $r$ and $\tau$ refer to the radial and tangential components.
% P = gen = [div =0 in hom] = sphr coords
% based on the second-order susceptibility tensor
%
% that it allows for the separation of the radiation
% generated from different multipoles of the SH source.
The bulk polarization depends on the spatial derivatives (gradients) of the electromagnetic field % is recovered  by
\begin{multline}
\label{Pbulk}
{\bf{P}}_{\text{bulk}}^{^{\left( {2\omega } \right)}} =
\; \beta {{\bf{E}}^{\left( \omega  \right)}}\nabla  \cdot {{\bf{E}}^{\left( \omega  \right)}} +
   \; \gamma \nabla \left( {{{\bf{E}}^{\left( \omega  \right)}} \cdot {{\bf{E}}^{\left( \omega  \right)}}} \right) \\ +
\; \delta' \left( {{{\bf{E}}^{\left( \omega  \right)}} \cdot \nabla } \right){{\bf{E}}^{\left( \omega  \right)}}
% \text{~~~~~~~~~~~~~~~~~~in } \Omega_i
\text{ ,}
\end{multline}
where $\beta$, $\gamma$ and $\delta'$ are material parameters~\cite{Heinz,Dadap2004}.
In the homogeneous media, the first term vanishes. The component $\chi^{\left( {2} \right)}_{ \bot  \bot  \bot }$ of the surface susceptibility tensor commonly yields the largest contribution to the surface SH response for metal nanopartciles~\cite{Wang2009,Bachelier2010} and reaches the values of $\sim 10^{-18}$ m$^2$V$^{-1}$~\cite{Krause2004}.
% The last term can be neglected in case of damping.
Aside from phenomenology, relying on experimental retrieval of nonlinear susceptibilities, analytical expressions for parameteres, shedding light on the origin of complex electronic nonlinearity in metals, can be derived within free-electron, hydrodynamic and quantum density functional theories~\cite{Sipe1982,Liebsch1988,Cirac2012,Sundararaman2014}.
A more involved self-consistent nonperturbative description of harmonic generation in metallic
nanostructures was recently developed through transient implementation of the hydrodynamic model~\cite{Ginzburg2015,Krasavin2016}.

\begin{figure*}[t!]
% \centerline{\mbox{\resizebox{8.4cm}{!}{\includegraphics{Fig2_Plasmonics_v1}}}}
\centering\includegraphics[width=0.7\linewidth] {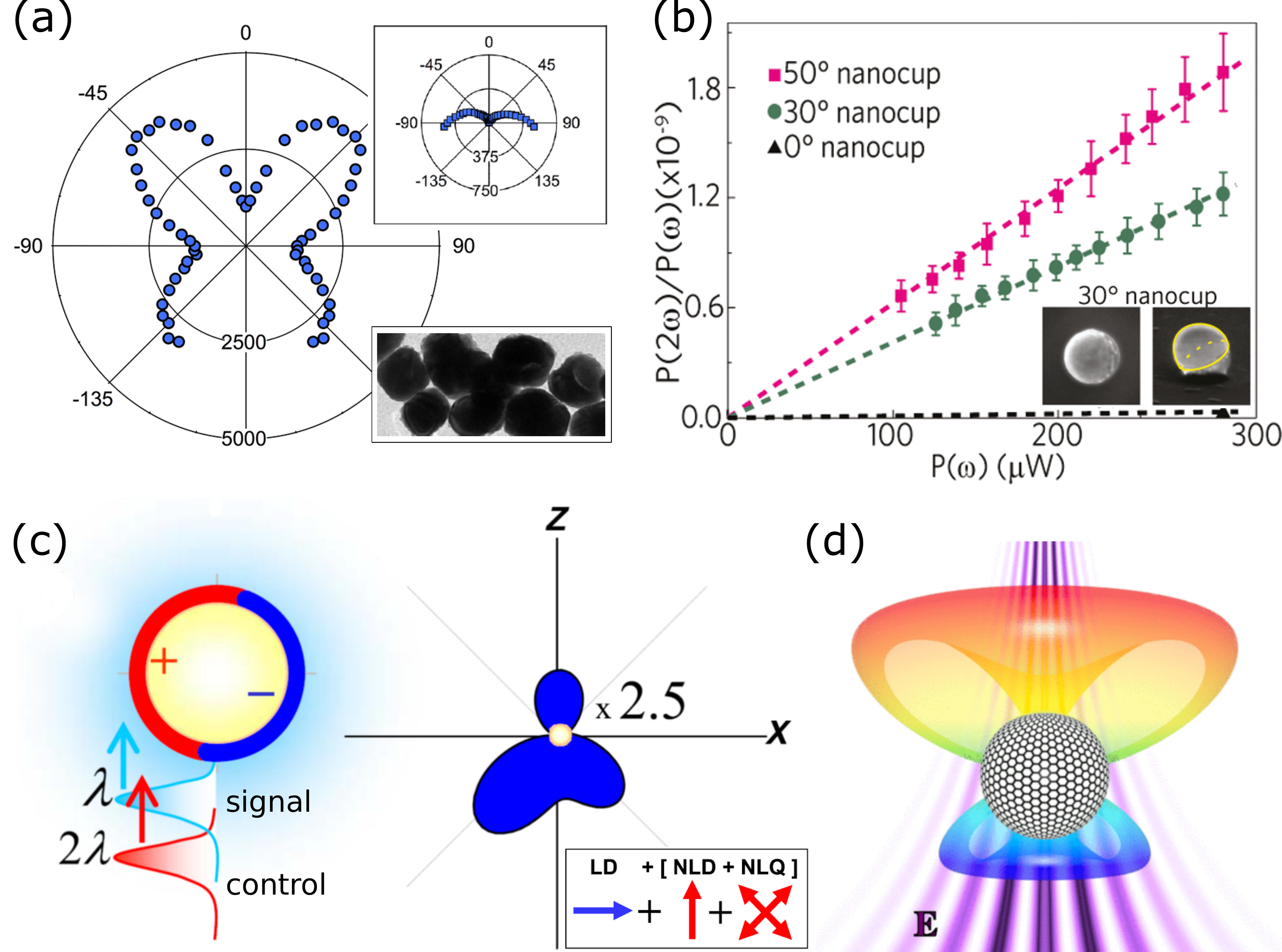}
\caption{(a) SH scattering pattern from Ag nanoparticles in water with four well-resolved quadrupolar lobes.
% lobes evidencing a quadrupolar .
Insets: SH scattering pattern from MG/PS nanoparticles featuring a dominant nonlocally excited electric dipole (top); TEM image of Ag nanoparticles (bottom). (b) SH conversion efficiency by individual nanocup scales linearly as a function of the incident laser power and increases with nanocup angle. Inset: SEM image of a chemically synthesized nanocup on a glass substrate oriented at $30^{\circ}$ to the substrate normal. %35nm-thick Au layer is deposited by electron-beam evaporation on
120 nm silica nanoparticle is capped with a 35nm-thick Au layer. (c) Coherent control of light scattering from a gold nanowire by two collinear signal and control beams with zero phase delay. (d) Schematic view of a graphene-wrapped
nanoparticle placed into an axially symmetric slightly
inhomogeneous external field. The SH radiation
predominantly directed into the upper half-space.
Panels (a-d) are adopted from Refs.~\cite{Gonella2012},~\cite{Zhang2011},~\cite{Rodrigo2013},~\cite{Smirnova2014}, respectively.
}
\label{fig:fig2}
\end{figure*}
%

%-----------------------------------------------------------------------------------------------------------------------------------
% ANALYTICAL APPROACHES: NLM and NLRGD
% NONLINEAR MIE THEORY SPHERE~\cite{Dadap1999,Dadap2004,Pavlyukh2004,deBeer2009,Jen2010,Gonella2011,Wunderlich2011,Butet2012}
% ARBITRARY SHAPED~\cite{deBeer2011}, ~\cite{deBeer2007}
%
% inspect explore
To describe the nonlinear scattering theoretically, the nonlinear Mie theory %(NLM)
was developed over recent decades~\cite{Dewitz1996,Pavlyukh2004,deBeer2009,Gonella2011,Wunderlich2011,Butet2012,Roke2012,Capretti2013} as an extension of the analytical approach outlined in Refs.~\cite{Dadap1999,Dadap2004} for the Rayleigh limit of SHG from a spherical particle. This theory can be used for particles of any sizes and materials, however being limited to strictly spherical shapes, and it is widely exploited to examine SHG governed by the dominant surface SH polarization source~\cite{Bachelier2010,Wang2009} in spheroidal metallic nanoparticles. In contrast, another % analytical method,
technique, the nonlinear Rayleigh-Gans-Debye model is applicable for particles of any shapes as long as a refractive index mismatch between the particle and a host medium is low. However, the latter is not the case for metallic particles immersed in aqueous medium, frequently employed in experiments.
% Computational %()
Numerical studies in nonlinear nanophotonics~\cite{Biris2010,Biris2010_oe,Gallinet2015} are most commonly %(based on)
performed with the use of finite-element~\cite{Bachelier2008,Butet2010PRL,Butet2010,Butet2015josab} method, integral equation techniques~\cite{Forestiere2013,Butet2013,Butet2014_2}, hydrodynamic models~\cite{Cirac2012,Ginzburg2015} and FDTD~\cite{Laroche2005}.
% spherical particles clusters theory~\cite{Butet2014_2}

In the small-particle limit $k R \ll 1$, % $k_0 R \ll 1$
it was proven analytically that nonlinear local electric quadrupole and nonlocal electric dipole, 
originating from the retardation effect, excited in a spherical metal nanoparticle of radius $R$ at the
SH frequency provide the main contributions to
the SHG radiation~\cite{Dadap1999,Dadap2004}, whereas
the linear scattering properties of small particles are largely determined by the electric dipolar response.
This deduction has been also supported %(evidenced)
by extensive experimental studies~\cite{Shan2006,Jen2009,Butet2010}.
% Additionally,
Deformation effects, causing the deviation of the particle shape from a perfect sphere, allow for the SH emission from the locally excited dipole~\cite{Nappa2005,RussierAntoine2007}.
With increasing the particle size, the onset of an octupolar contribution was detected % observed
for Au nanoparticles of $70$ nm in diameter~\cite{Butet2010PRL,Bachelier2010}.
% Early experiments BUTET, Later - octupole, Qaudrupole for sensitivity,
% Size dependence is demonstrated Gonella in figure for,

Figure~{\ref{fig:fig2}}(a) demonstrates the angle-resolved second-harmonic scattering (AR-SHS) pattern from 80 nm diameter Ag nanoparticles in water excited at the fundamental wavelength $800$ nm~\cite{Gonella2012}. In agreement with theoretical predictions, the dominant surface susceptibility component $\chi^{\left( {2} \right)}_{ \bot  \bot  \bot }$ drives a qaudrupolar-like SH emission. For comparison, AR-SHS pattern from nonplasmonic similarly-sized spherical polysterene nanoparticles adsorbed with nonlinear-optically active malachite green molecules (MG/PS) (88 nm diamter) excited at $840$ nm is shown in the inset. In the latter case, nonlocally excited dipole emission characterized by two lobes stems from the different leading component $\chi^{\left( {2} \right)}_{ \bot \parallel \parallel}$.

%-----------------------------------------------------------------------------------------------------------------------------------
% structure with reduced symmetry
%
% SHG supported stymmetries cand selection rules
% mirror symmetries of a scatter
Spectral and angular properties of SHG from plasmonic nanoparticles are largely influenced by structural symmetries of both scatterer and external field,
a type of excitation~\cite{Mochn2003,Huo2011}, % , % e. g. plane wave versus focused beam~\cite{Huo2011}
spatial variations of the applied electromagnetic field % ~\cite{Mochn2003}  % field inhom~\cite{Mochn2003},
and field gradients at nanoscale, and it can be significantly enhanced in the vicinity of plasmonic resonances known for their inherent geometric tunability.
For instance, synthetic nonlinear media can be designed from reduced-symmetry 3D plasmonic elements, % particles components building blocks constituent
such as nanocups consisting of a spherical metal semishell fabricated around a dielectric nanoparticle core~\cite{Zhang2011}.
Symmetry breaking in this geometry introduces a magnetic dipole mode, when
a nanocup resembles a resonant LC circuit.  % For this mode, a nanocup resembles a resonant circuit.  where functions
Both intensity and emission profiles of the enhanced SHG from such a nanocup excited at the magnetic dipole resonance were observed to depend on a nanocup orientation. The conversion efficiency increases dramatically as the angle between the pump excitation direction
and the nanocup symmetry axis is increased, as shown in Fig.~{\ref{fig:fig2}}(b), while the scattering directionality is dictated by a nanocup inclination, regardless of the excitation direction or polarization.
%
% form variation, broken structural symmetry of particles, field inhomogeneity, plasmonic resonances

%-----------------------------------------------------------------------------------------------------------------------------------
% NOVOTNY antenna~\cite{Rodrigo2013}, coherent control more THG ~\cite{Zeuner2015}
In the context of plasmonic nanoantennas, the idea of all-optical control of the radiation directionality
through an interplay between linear scattering
and second-harmonic generation was proposed in Ref.~\cite{Rodrigo2013}.
In the proof-of-concept theoretical model, the authors consider a 100 nm gold nanowire % in air
simulataneously excited by a low-intensity (signal) and high-intensity (control)
fields %(waves)
at wavelengths $\lambda = 532$ nm and $2\lambda$, repectively.
Asymmetric scattering pattern occurs owing to the interference between the linear electric dipole (LD) induced by the signal and nonlinear dipole (NLD) and quadrupoles (NLQ) generated by the control, as exemplified in Fig.~{\ref{fig:fig2}}(c). % With varying
Altering excitation conditions, the authors demonstrate that light scattering from a single nanoelement can be efficiently tuned by changing phase delay and relative angle between two excitation beams.
Similarly, in Ref.~\cite{Zeuner2015} it was shown that THG signal emitted from a three-nanorod gold structure can be coherently controlled by changing a relative phase of two different excitation channels. Localized surface plasmon polaritons in the nanorods are excited through near- and far-field couplings to two orthogonally polarized light fields. The cancellation of the excitation was measured as a minimum of the THG signal, and the observed phase shift
between the THG signal and excitation was attributed to damping effects.

%-----------------------------------------------------------------------------------------------------------------------------------
% GRAPHENE
Graphene, a single atomic layer of graphite, has emerged recently as a promising alternative
to noble metals for applications in plasmonics~\cite{RevGrigorenko,9}.
The study of plasmonic effects
in doped graphene structures has attracted a special interest from the nanoplasmonics research
community due to novel functionalities suggested by such systems, including an
extraordinary field confinement by a graphene layer, tunability of graphene properties
through doping or electrostatic gating and longer lifetimes in the infrared and terahertz
frequency ranges~\cite{JablanReview}.
%, which is extremely important for biomedical and security applications.
In addition, graphene demonstrates strong and tunable optical nonlinearity and it can be
incorporated into various components of nanoscale optics~\cite{Glazov2013,Cox2014,Manzoni2015}.
The study of the nonlinear response of graphene in the resonant plasmonic geometries is still in its infancy and is expected to bring new concepts and find real applications. % and is expected to
%
% We develop theoretical models for the resonant (enhanced) second-harmonic generation
% from a graphene-wrapped dielectric spherical nanoparticle
%
Remarkably, graphene is not only employed for planar geometries,
but there are also attempts to wrap complex three dimensional
objects in graphene~\cite{GrWrappingKo, GrWrappingLee, GrWrappingSheng}.
In Ref.~\cite{Smirnova2014}, a theoretical model for the resonant second-harmonic generation
from a graphene-wrapped dielectric spherical nanoparticle has been developed.
Strong nonlinear response is caused by an induced quadratic surface current in graphene.
When the particle is illuminated by a plane wave, the resonantly enhanced radiation
of the second harmonic is quadrupolar and symmetric, however, the radiation pattern of the second harmonic
can be manipulated by placing the particle in a weakly inhomogeneous
external field, which is schematically shown in Fig.~{\ref{fig:fig2}}(d).
This resultant asymmetric emission is determined by the interference of dipole and quadrupole modes and
the asymmetry of the external field changes the efficiency of the excitation
of these modes.  For the typical parameters,
the second harmonic is predominantly radiated in the halfspace
with stronger external electric field.

%-----------------------------------------------------------------------------------------------------------------------------------
% KAURANEN experiment nonspherical L-shape particles~\cite{Kujala2007}, chi2 tensor analysis ~\cite{Kujala2008, Zdanowicz2011, Huttunen2012}, noncentrosymmetric metal nanoparticles (metamolecules) L-shaped~\cite{Husu2012}, T-shape~\cite{Canfield2007} field gradients at nanoscale
A macroscopic description of the nonlinear response from arrays of noncentrosymmetric nanoparticles is based on the effective multipolar tensors~\cite{Canfield2006,Kujala2008, Zdanowicz2011, Huttunen2012}.
In particular, the magnetic dipole and electric quadrupole contributions were experimentally identified % revealed
in the nonlinear emission of L-shape gold nanoparticles through asymmetry of the generated SH field intensity measured in transmission and reflection~\cite{Kujala2007}.
The second-order nonlinear response of arrays of metal nanostructures and the relative importance of such multipolar effects %, in the scattering-matrix formalism and tensor,
are generally affected by
the symmetry properties and shapes of scatterers, quality of samples, plasmonic enhanement and small-scale inhomogeneities of the local field, hot spots,
particle %ordering (mutual 
arrangement and interparticle coupling~\cite{Canfield2006OE,Canfield2007,Husu2008,Husu2012}.  
% long-range diffrcative coupling between the particles
% plasmonic enhanement and small-scale inhomogeneities of the local fields
% higher-multipolar interactions
%

Being the simplest nonlinear optical process, SHG was intensively studied in
% a plethora of
various plasmonic structures of different shapes and coupling~\cite{Klein2006,Canfield2007,Husu2012,AvanderVeen2015,Walsh2013,Black2015}, % including , to name a few.
chiral~\cite{Valev2010,Valev2011,Valev2014} and Fano-resonant geometries~\cite{Thyagarajan2013,Liu2016}.
Based on near- and far-field radiation properties, % concrete unsophisticated 
specific designs have been suggested for applications in sensing~\cite{Butet2012,Gonella2014}, shape chracterization~\cite{Butet2013_2}, nonlinear nanorulers~\cite{Butet2014,Shen2015} and nonlinear microscopy~\cite{Shen2013}.

\section*{Metal-dielectric structures}

\begin{figure*}[t!]
% \centerline{\mbox{\resizebox{8.4cm}{!}{\includegraphics{Fig2_Plasmonics_v1}}}}
\centering\includegraphics[width=0.97\linewidth] {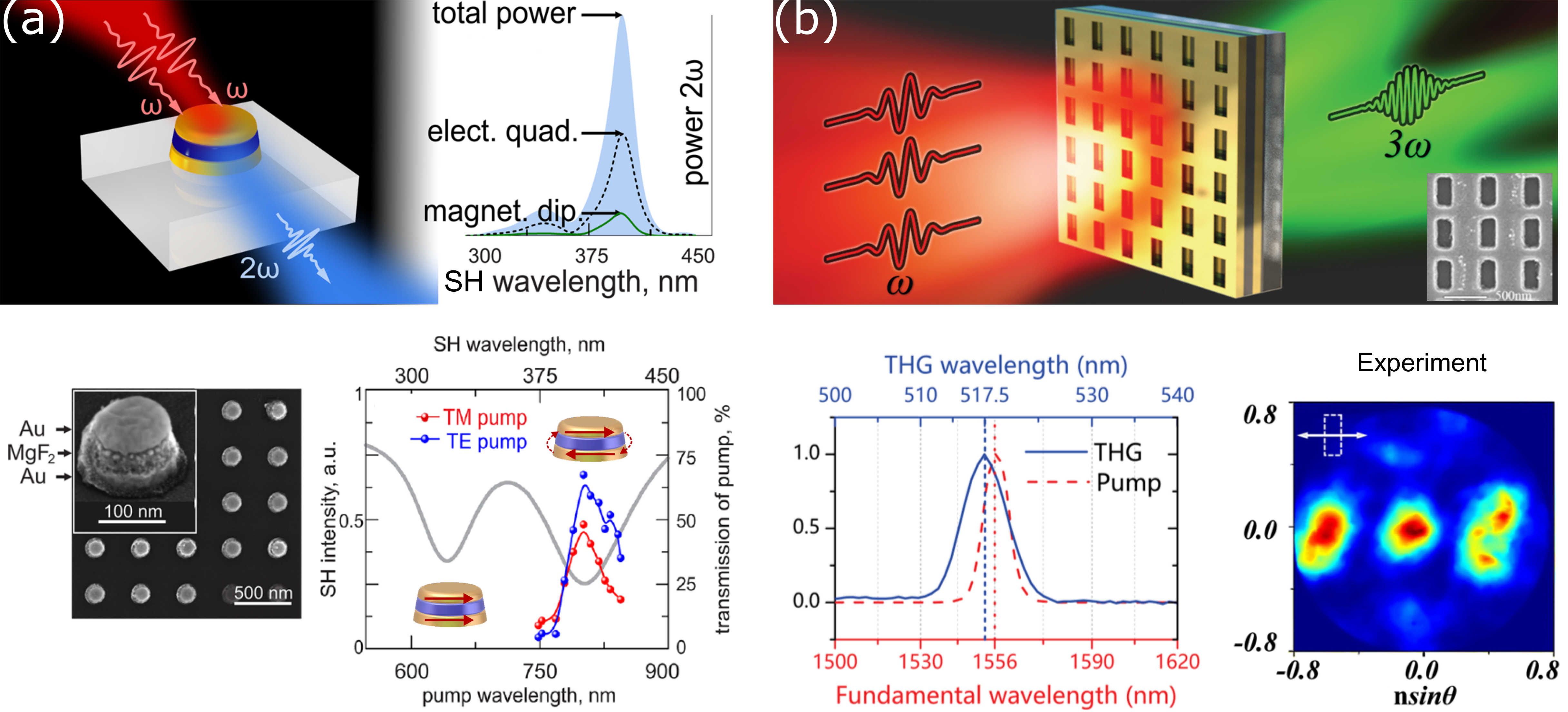}
\caption{
% Silicon-based nonlinear nanoparticles and metasurfaces.
(a)
Top left: Schematic view of the experiment on SHG with a metasurface of metal-dielectric disk-like nanoparticles made of Au/MgF2/Au
layers. Top right: Multipole decomposition of SH signal numerically calculated for the TE-polarized pump at the
oblique 30$^{\circ}$ incidence. 
Bottom left: SEM image of a fabricated metasurface. Bottom right: Linear transmission of the
metasurface (gray) at normal incidence and second-harmonic spectra
for TM- and TE-polarized pump waves (red and blue) measured for
the incident angle of 30$^{\circ}$.
Adopted from Ref.~\cite{Kruk2015}. 
The insets illustrate the electric current distribution in a single composite nanoparticle in ED and MD resonant modes associated with two dips in transmission. 
(b) Top: Schematic of the experiment on the resonantly enhanced THG from a fishnet metamaterial, exhibiting a magnetic response. The inset shows a scanning electron micrograph image of the sample. 
Bottom left: Measured THG (solid blue) and pump (dashed red) spectra, normalized to maxima. Bottom right: back-focal-plane Fourier image of TH radiation.  
Adopted from Ref.~\cite{Wang2016}.
}
\label{fig:figMD}
\end{figure*}

Recently, it was established that very specific nonlinear regimes can be achieved due to magnetic optical response of artificial "meta-atoms"~\cite{Shadrivov2015_book}.
This picture is at large adapted in the microwave range, and a vast number of research groups are engaged in utilizing tailored nonlinear response of metamaterials, as shown in the recent review~\cite{Lapine2014}.
When nonlinearities of both electric and magnetic origin are present,
nonlinear response can be modified substabtially being accompanied by nonlinear interference,
magnetoelectric coupling, and wave mixing effects~\cite{Rose2013, Rose2013_2}. In particular, the interference between
the electric and magnetic nonlinear contributions can lead to unidirectional harmonic generation from an ultra-thin metasurface~\cite{Rose2013} and "non-reciprocal" nonlinear scattering~\cite{Rose2013} from a symmetric subwavelength nanoelement~\cite{Poutrina2016}.
The idea of magnetic nonlinearities evolves, being further elaborated
by the optics community. It finds % realizations
implementations in double-layer fishnet structures~\cite{Kim2008,Reinhold2012,Shorokhov2016_2,Wang2016}, % nanoburgers,
coupled metal nanodisks~\cite{Kruk2015,Kolmychek2015}, and nanostrips~\cite{Chandrasekar2015}.
These coupled plasmonic structures sustain antisymmetric oscillations of electron plasma leading to the magnetic-type resonances and the associated enhancement of the nonlinear response.
The dissimilar electric and magnetic contributions
to the nonlinear response are then disclosed by % performing
spectrally resolved measurements and multipolar decomposition of the generated harmonic signal. 
In Ref.~\cite{Kruk2015}, such analysis was performed for SHG observed from a metasurface composed of metal-dielectric-metal nanodisks, see Fig.~{\ref{fig:figMD}}(a). 
The resonant excitation of the magnetic optical mode in composite nanoparticles was shown to govern the efficient nonlinear conversion of the multipolar origin, being a superposition of nonlinearly-induced magnetic dipole and electic quadrupole modes with a suppressed contribution from the electric dipole mode. 
Phenomenological macroscopic description, developed in Ref.~\cite{Kolmychek2015}, established a connection of the resonant enhancement of SHG in such "nanosandwiches" under the excitation of the MD resonance with the dominant role of the nonlinear magnetic-dipole polarization driven by the $\chi^{emm}$ susceptibility. Here, the first index in the superscript corresponds to the electric polarization induced at SH frequency, while the two other denote the magnetic fundamental fields.   
 
Ref.~\cite{Wang2016} reports on the experimental identification % (analysis)  multipolar origin  
of multipolar contributions in the THG signal from a metal-dielectric-metal layered fishnet metamaterial illuminated with a focused pump beam in the vicinity of the magnetic resonance, see Fig.~{\ref{fig:figMD}}(b). % originating from 
Previously, the electric and magnetic dipole resonances in fishnets were shown to be reponsible for notably distinct THG angular patterns~\cite{Reinhold2012}. 
The metamaterial studied in Ref.~\cite{Wang2016} consists of Au/MgF$_2$/Au structure perforated periodically with an array of holes, and the magnetic response is attained due to antiparallel currents in the top and bottom metal layers. 
The detected THG radiation pattern, measured by a Fourier imaging technique, was proven % shown
to be a result of the interference of the electric dipolar, magnetic dipolar and electric quadrupolar modes~\cite{Wang2016}. 

\section*{All-dielectric nanostructures}

Ohmic losses, heating and low damage thresholds % and melting temperature
pose substantial restrictions on the  achievable performance of the nonlinear plasmonic devices exploiting metallic components, thus limiting their practical use.
Nanoparticles made of high-refractive-index dielectrics and semiconductors (such as germanium,  % ($n\sim 3.5$)  ($n\sim 4$) $n\sim 4$  n=3.4
tellurium, GaAs, AlGaAs, GaP, silicon), which do not suffer from large intrinsic absorption at the visible, infrared and telecom frequencies, are emerging as a promising alternative to plasmonic nanostructures for nanophotonic applications~\cite{Albella2014,Jahani2016}.
Such high-permittivity nanoparticles exhibit strong interaction with light due to the excitation of the
Mie-type resonances they sustain~\cite{GarcaEtxarri2011,Krasnok2011ask,Krasnok2012ask,Schmidt2012}.
Both these electric and magnetic resonances can be spectrally controlled and engineered independently, and
therefore all-dielectric nanostructures offer unique opportunities for the study of nonlinear effects owing
to low losses in combination with multipolar optical response
of both electric and magnetic nature.
	 Utilizing the Mie-type resonances in resonant subwavelength dielectric % nanospheres, nanodisks and others
	 geometries has been recently recongnized a promising strategy to gain
	 high efficiencies of nonlinear processes at low mode volumes and design novel functionalities originating from the
	 optically-induced magnetic response~\cite{Shcherbakov2014,Sanatinia2014,Shcherbakov2015,Shcherbakov2015_2,Yang2015,Carletti2015, Maragkou2015,Makarov2015,Carletti2016,thg_dasha,Dmitriev2016,Shorokhov2016,Baranov2016, Grinblat2016,Gili2016}.
Particularly, optically magnetic dielectric nanostructures
are shown to significantly enhance nonlinear conversion,
with the high enhancement observed for excitation of the magnetic dipole resonance. % highest

\begin{figure*}[t!]
% \centerline{\mbox{\resizebox{8.4cm}{!}{\includegraphics{Fig2_Plasmonics_v1}}}}
\centering\includegraphics[width=0.97\linewidth] {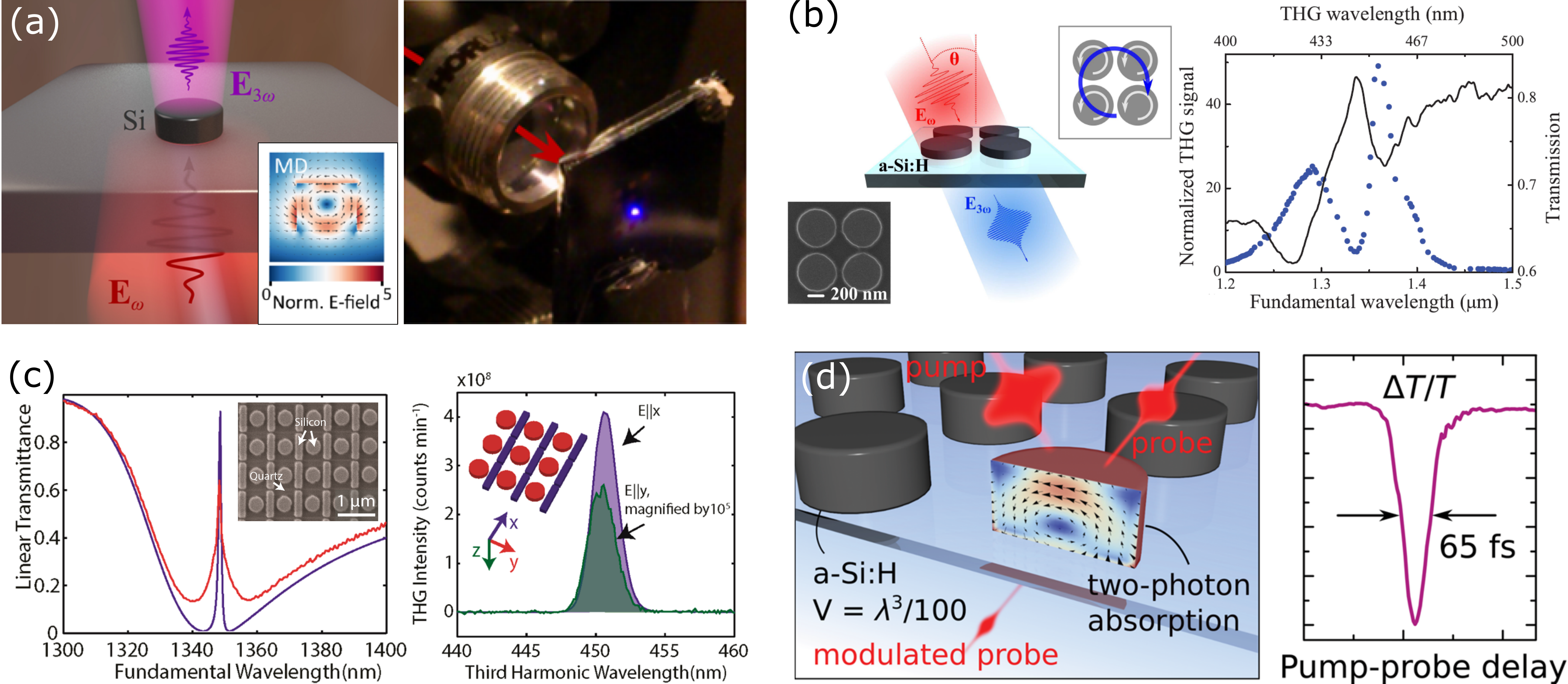}
\caption{
% Silicon-based nonlinear nanoparticles and metasurfaces.
(a)
Left: Illustration of THG from silicon nanodisks driven by the magnetic dipolar resonance excited in the disks by the impinging laser light at the fundamental frequency. %  optical
Superimposed is exemplery near-field distribution % profile
of the induced electric field in a disk at MD resonance.
Right: A photographic image of the eye-visible generated radiation from the nanodisks sample.
Adopted from Ref.~\cite{Shcherbakov2014}.
(b)  Left: Schematic illustration of the resonant THG from silicon quadrumers of four a-Si:H nanodisks with SEM image of the sample in the bottom left corner.
Rotating arrows picturize the origin of the magnetic Fano resonance in quadrumers: two coupled magnetic-like modes formed by out-of-plane magnetic dipoles and circulating displacement current produced by in-plane electric dipoles give rise to the Fano interference. For details, see Ref.~\cite{Hopkins2015}.
Right: THG spectroscopy of  a-Si:H quadrumers on a glass substrate. Shown are experimental transmission
(black line) and THG (blue dots) spectra of the sample excited by obliquely incident s-polarized radiation at the angle of 45$^{\circ}$.
TH power is plotted normalized to the spectrum of a bare a-Si:H film.
After Ref.~\cite{Shorokhov2016}.
(c) Left: Simulated (blue) and experimentally measured (red) transmission
spectra of the silicon Fano-resonant metasurface with the Fano peak at the wavelength of 1350 nm. SEM image of the fabricated structure is shown in the inset.
Right: Third harmonic spectra of the metasurface excited with the incident electric field polarized along the long ($x$-) and short ($y$-) axes of the bars, respectively. Adopted from Ref.~\cite{Yang2015}.
(d) Ultrafast all-optical switching in resonant silicon nanodisks:
(left) schematic of the experiment: a-Si disc 250 nm in diameter is capable of switching optical pulses at femtosecond rates;
(right) measured modulation of the probe pulse transmittance as a function of time delay between the probe and pump pulses at
low pump powers. Taken from Ref.~\cite{Shcherbakov2015_2}.
}
\label{fig:fig3}
\end{figure*}

Remarkably, the observation of highly localized fundamental magnetic dipolar mode characterized by circular displacement currents
excited in spherical silicon particles of sub-micrometer size in the visible spectrum~\cite{Evlyukhin2012,Kuznetsov2012} has led to the development of "magnetic light" concept, which triggers the whole research direction of {\em all-dielectric nanophotonics}~\cite{Ginn2012,Jahani2016}.
In most previous works on trapped magnetic resonances,
silicon was the primary focus of the material~\cite{Evlyukhin2012,Kuznetsov2012,Fu2013,Staude2013,Zywietz2014,Liu2014,Zywietz2015} thanks to its CMOS compatibility, low cost and well-established fabrication methods.
Along with a moderately high refractive index, silicon, both crystalline and amorphous, being a centrosymmetric material with inhibited quadratic nonlinearity, % second-order
possesses however a strong bulk third-order optical susceptibility % $\chi^{(3)_{b}}$
$\mathord{\buildrel{\lower3pt\hbox{$\scriptscriptstyle\leftrightarrow$}} \over \chi }^{(3)}_b$~\cite{Burns1971,Moss1989,Moss1990,Bristow2007,Lin2007,Zhang2007,Ikeda2007,VivienBook,Gai2014}, nearly 200 larger than that of silica; therefore strong nonlinear phenomena can be anticipated in the optical response of
silicon nanoparticles. 
% Below we review the recent investigations of nonlinear optical properties of silicon-based
% nanoparticles and metasurfaces, where the nonlinearity
% is enhanced or controlled using the unit cell geometry.
Figures~{\ref{fig:fig3}}(a--d) % summarizes the recent investigations 
present a few examples of the recent experimental study of nonlinear optical properties of silicon-based nanoparticles, their clusters, and metasurfaces.  

{\em Experimental studies of third-harmonic generation.} In the pioneering work~\cite{Shcherbakov2014}, it was experimentally confirmed that a strong enhancement of THG can be achieved in isolated silicon nanodisks and their
arrays optically pumped in the vicinity of the magnetic dipolar resonance by using THG microscopy and THG spectroscopy techniques.
The MD resonance is characterized by the significantly enhanced local fields tightly bound within the volume of nanoscale resonators.
Since the induced volume nonlinear source, third-order polarization ${\bf P}^{(3\omega)}_{\text{bulk}}$, scales with the local electric field ${\bf E}^{(\omega)}_{\text{loc}}$ cubed,
% ${\bf P}^{(3\omega)}= \chi^{(3)} {\bf E}^{(\omega)}_{\text{loc}}$, where $\chi^{(3)}$ is the third-order susceptibility of silicon, and ${\bf E}^{(\omega)}_{\text{loc}}$
% is the local complex electric field at the fundamental frequency.
$ {\bf P}^{(3\omega)}_{\text{bulk}}= \varepsilon_0 \mathord{\buildrel{\lower3pt\hbox{$\scriptscriptstyle\leftrightarrow$}} \over \chi }^{(3)}_b {\bf E}^{(\omega) 3}_{\text{loc}}$, a considerable enhancement of THG is expected from the nanodisks pumped by the resonant laser radiation.
In the experiment,
silicon nanodisks placed on silica were subject to an intense femtosecond laser pulse train with frequency close to the magnetic dipole resonance.				
Analyzing the transmitted signal at TH, it was found that the resonant THG response from silicon nanodisks prevails by the factor of up to 100 over the THG from the bulk silicon slab. The generated 420 nm radiation
was bright enough to be observed by a naked eye under the table-lamp illumination conditions, see Fig.~{\ref{fig:fig3}}(a).

%%% ACS Phot oligomers
Combining nanodisks into oligomers provides another degree of freedom in tailoring the nonlinear optical response.
Particularly, strong reshaping of the third-harmonic spectra was observed from trimers of silicon nanodisks
with varying interparticle spacings~\cite{Shcherbakov2015}.
Complex multipeak patterns in measured THG spectra
were substantiated by interference of the nonlinearly generated TH waves radiated by different parts of the sample and augmented by an interplay between the electric and the magnetic dipolar resonances in the nanodisks.

Optical nonlinearities can be further enhanced in finite nanoparticle clusters
through the excitation of high-quality collective modes of the magnetic nature.
Combining dielectric particles in oligomers may lead to the formation
of collective modes with different lifetimes limited by radiative losses mainly, in a striking contrast to deeply
subwavelength plasmonic analogues. Such multiresonant dielectric nanostructures can be designed to support high-quality modes of different symmetries,
amplifying nonlinear optical effects. In this context, Ref.~\cite{Shorokhov2016} presents the study of
THG from quadrumers composed of four identical silicon nanodisks.
Nontrivial wavelength and angular dependencies of the generated
harmonic signal were observed, featuring a high enhancement
of the nonlinear response associated with
a specific type of the magnetic Fano resonance in oligomeric dielectric systems. 
First predicted in Ref.~\cite{Hopkins2015}, 
this Fano resonance arises due to the magneto-electric
dipole coupling in individual quadrumers. The measured linear transmission
displayed in Fig.~{\ref{fig:fig3}}(b) shows a broad transmission dip and a sharp
Fano-like line shape at longer wavelengths. They are attributed to the dominant
resonant excitation of different collective modes in the quadrumers.
Both resonances are characterized by pronounced magnetic dipolar excitations in the disks and
significant local field enhancement and thereby they are accompanied by a substantial increase of the third-harmonic signal,
as compared to that from both optically-decoupled disks and an unstructured silicon film of the same thickness.

Noticeably, such mechanism for the THG enhancement in a single unit cell of all-dielectric metasurfaces is different from that reported in Ref.~\cite{Yang2015}, where
the efficient THG is driven by a high-quality resonance, achieved in a two-dimensional array of coupled silicon disks and bars,
which stems from collective bright and dark modes of the lattice [see Fig.~{\ref{fig:fig3}}(c)]. In the latter design, nanobars with long axes parallel to the E-polarization of a normally incident radiation
and exhibiting electric dipole resonances effectively couple energy from the incident
wave into the dark out-of-plane magnetic dipole modes of neighboring silicon disks.
A sharp peak in the measured transmittance spectra indicates the Fano interference featuring a resonance with an experimental $Q$-factor of 466.
The associated strong enhancement of the local electric field trapped within the disks gives rise to
to THG amplification factor of $1.5 \times 10^5$ with respect to the unpatterned silicon film, leading to the conversion efficiency on the order of $10^{-6}$, which is the largest value reported for THG on nanoscale using comparable pump energies~\cite{Yang2015,Grinblat2016}. 

A simple disk geometry of dielectric nanoparticles appears especially attractable for experimental studies of resonant nonlinear effects 
because the positions of both their electric and magnetic Mie-type resonances can be easily tuned by varying the
aspect ratio~\cite{vandeGroep2013}. 
For instance, the efficient excitation of the magnetic dipole mode requires sufficient 
field retardation throughout the nanodisk to drive the displacement current loop. 
This suggests that that if the particle gets shallow enough, ED and MD resonances become overlapped gradually expelling the in-plane MD mode.   
% and therefore a MD resonance cannot be supported
Ref.~\cite{Grinblat2016} demonstrates that THG in thin Ge nanodisks under normally incident laser excitation can be boosted via the nonradiative anapole mode (AM). 
This AM is accompanied by the pronounced dip in scattering occurring from destructive far-field interference
of the electric and toroidal dipole moments, and its physics can be inferred from the Fano resonance mechanism~\cite{Miroshnichenko2015}. % explained 
The AM, which emerges for low enough aspect ratios, efficiently confines the electric energy inside the disks, which is similar to the performance of MD mode in thicker disks. The TH conversion at the anapole mode from a germanium disks array was measured to outperform that at the radiative electric dipolar modes, spectrally surrounding AM, by about one order of magnitude.

\begin{figure*}[t!]
% \centerline{\mbox{\resizebox{8.4cm}{!}{\includegraphics{Fig2_Plasmonics_v1}}}}
 \centering\includegraphics[width=0.65\linewidth] {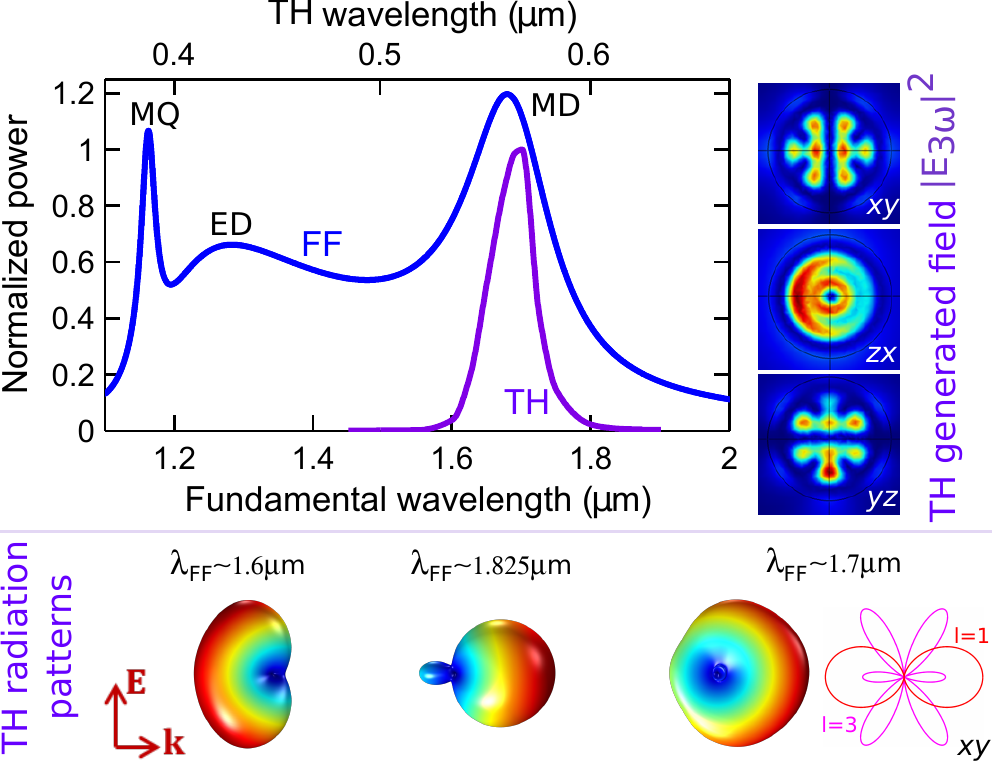}
\caption{Spectra of the normalized scattered FF (blue) and radiated TH (purple) powers calculated numerically for a spherical dielectric particle of radius $230$ nm and refractive index $3.5$ excited by a linearly polarized plane wave with electric field directed along the $x$ axis. Labeled are the positions of Mie resonances (MQ, ED and MD) in the linear scattering.
% over the depicted wavelength range. %marked
Intensity maps show spatial distributions %(maps) 
of the TH field generated near MD resonance in different cross-sections. 
Bottom: transformations of the TH radiation pattern with switching directionality. 
Within the broad MD resonance, both magnetic dipole and magnetic
octupole contribute to the THG peak, that determines predominantly axially symmetric emission profile (right) and a six-petalled TH near-field structure. 
The $xy$ cuts of radiation patterns for the corresponding pure magnetic dipole, $l=1$ (red) and magnetic octupole, $l=3$ (pink) are depicted in the polar plot. %for reference.
Images are adopted from Ref.~\cite{thg_dasha}.
}  
\label{fig:figTHGsphere}
\end{figure*}

{\em Theoretical approach.}
Analytical and numerical analysis of the resonant THG from high-index dielectric nanoparticles was performed in Ref.~\cite{thg_dasha}.
This work discusses the basic features and multipolar nature of the nonlinear scattering near the Mie-type optical resonances from simple nanoparticle
geometries, such as spheres and disks, with a focus on silicon.
Considering a model of a high-permittivity spherical dielectric particle excited in the vicinity of
the magnetic dipole resonance, analytical expressions for the electromagnetic field generated at the tripled frequency are derived, and it is shown that the TH radiation pattern
can be treated as a result of interference of a magnetic dipole and magnetic octupole,
which is also confirmed by means of full-wave numerical simulations carried out in COMSOL, see Fig.~{\ref{fig:figTHGsphere}}. 
% Third-harmonic generation by a spherical high-index nanoparticle near the
% magnetic resonance.
Besides, in compliance with experimental findings~\cite{Shcherbakov2014,Shcherbakov2015,Shcherbakov2015_2,Yang2015}, the efficiency of THG at the MD resonance is noted to significantly exceed that at the ED resonance. Remarkably, due to a larger mode volume of the magnetic Mie mode compared to the electric mode of the same order, the magnetic dipole resonance also superiorly enhances Raman scattering by individual silicon nanospheres~\cite{Dmitriev2016}.
In addition, the approaches for manipulating and directing the resonantly enhanced nonlinear
emission by changing shapes and combining dielectic nanoparticles in oligomers are demonstrated.
This concept is illustrated with an example of a nanoparticle trimer
composed of a symmetric dimer of identical silicon spheres placed nearby a smaller nanoparticle.
Interparticle interaction influences the multipolar content
of the field scattered by a trimer, and the resultant constructive interference between the excited multipoles may provide high directivity of both the linear and nonlinear scattering simultaneously. Alternatively, the preferential backward/forward directionality of THG can be
attained for a silicon nanodisk of the adjusted aspect ratio exhibiting spectrally overlapped magnetic and electric dipole resonances.

{\em Ultrafast switching.} Silicon nanoparticles employed as nonlinear Mie-type cavities furthermore offer a platform for engineering compact and very speedy photonic switchers.
In Ref.~\cite{Shcherbakov2015_2}, the authors distinguish three basic classes of nonlinear phenomena responsible for the optical self-action effect in silicon: (i) Kerr-type
processes, including nonlinear refraction and two-photon absorption (TPA), (ii) free carriers (FC)
generation, and (iii) thermo-optical effect. They were able to design a metasurface of amorphous silicon nanodisks, illustrated in Fig.~{\ref{fig:fig3}}(d), where the undesirable free-carrier effects, which
are generally slow and pose serious restrictions on the speed of signal conversion, are suppressed, leaving the dominant fast TPA contribution in the nonlinear response.
This silicon-based metasurface allowed achieving strong self-modulation of femtosecond
pulses with a depth of 60$\%$ at picojoule-per-nanodisk pump energies.
Pump-probe measurements demonstrated that switching in the nanodisks
can be governed by pulse-limited 65 fs-long two-photon absorption
being enhanced by a factor of 80 with respect to the unstructured silicon film.
This result represents an important step towards novel and efficient active photonic devices, such as transistors
and logic elements to be built in integrated circuits~\cite{Maragkou2015}.
Such switching speeds potentially allow for creating data transmission and processing devices operating at tens and hundreds terabits per second.
Similar to Ref.~\cite{Shcherbakov2015_2}, Kerr and TPA effect-based transmission modulation was reported for a Fano-resonant metasurface in Ref.~\cite{Yang2015}.
Nonlinear light manipulation by scattering diagram and spectra of individual silicon nanoparticles also envisions
interesting possibilities for all-optical routing.
In Ref.~\cite{Makarov2015}, it was demonstrated, both theoretically and experimentally, that ultrafast transient modulation of the dielectric permittivity
due to variation of free-carrier (electron-hole plasma) density in silicon, induced via its photoexcitation by fs laser irradiation, significantly alters
scattering  behavior of a silicon nanoparticle, supporting a magnetic dipole resonance.
Later, accelerated reconfiguration of a radiation pattern from dipole-like to unidirectional (Huygens-source scattering) regime mediated by ultrafast (with times scales of about 2.5 ps) generation of electron-hole plasma during laser pulse-nanoantenna interaction was described in Ref.~\cite{Baranov2016}.
 
\begin{figure}[t!]
% \centerline{\mbox{\resizebox{8.4cm}{!}{\includegraphics{Fig2_Plasmonics_v1}}}}
 \centering\includegraphics[width=0.97\linewidth] {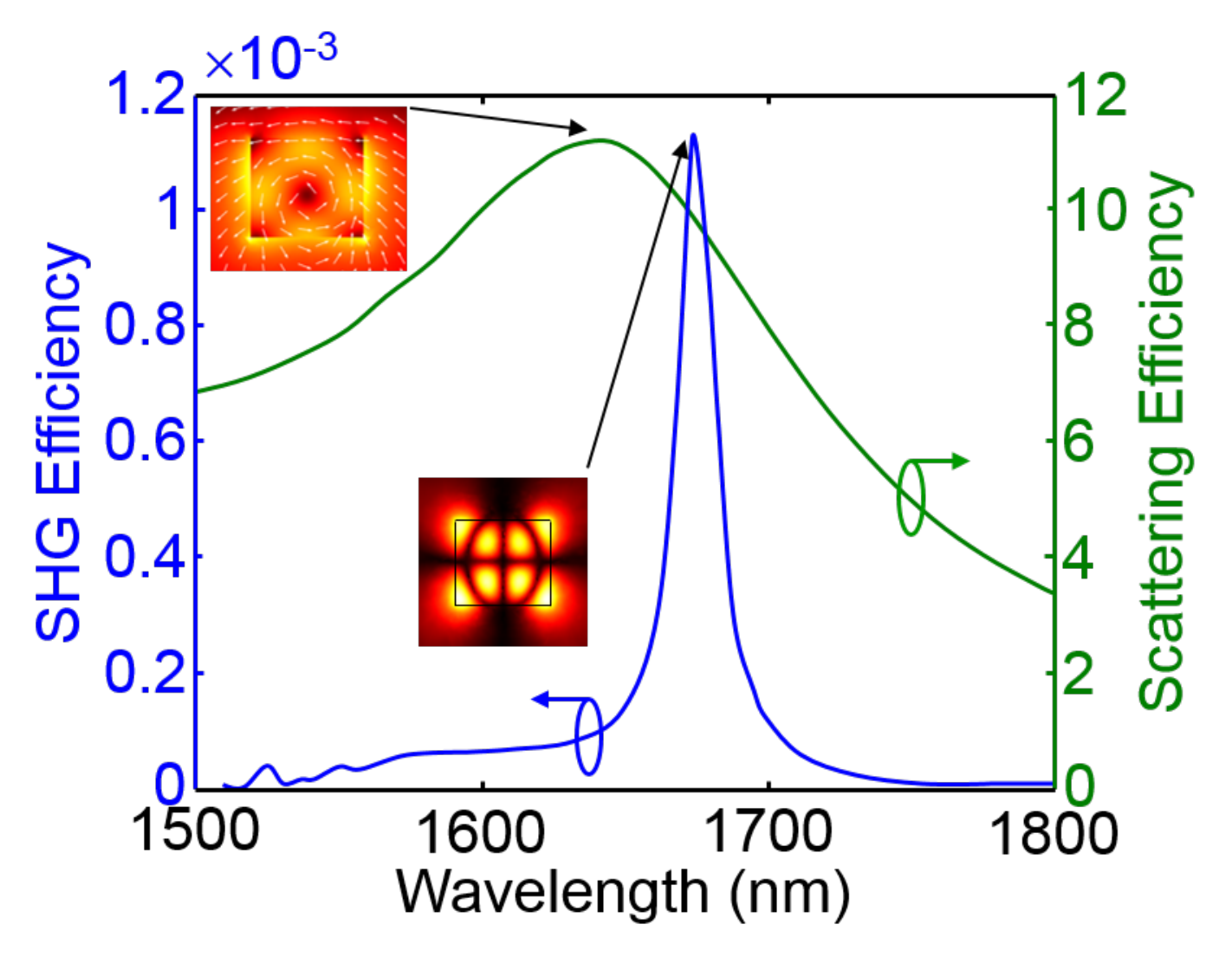}
\caption{SHG (blue line) and linear scattering (green line) efficiencies as functions of the pump wavelength calculated for a cylindrical AlGaAs nanoantenna % nanodisk
with a radius of $225$ nm and a height $400$ nm in air. %  suspended in air.  % cylinder
Insets show profiles of the electric field amplitude in a E-plane cross-section cut through the center of the cylinder, which correspond to the maxima of two dependencies dominated by MD and EQ modes, respectively. Adopted from Refs.~\cite{Carletti2015,Carletti2016}.
}
\label{fig:fig4}
\end{figure}

{\em Second-harmonic generation.} For efficient second-order nonlinear applications, advantageous can be excitation of low-order Mie modes in nanoantennas made of high-permittivity noncentrosymmetric semiconductors.
While in plasmonic nanoparticles the second-order nonlinear response is dominated by surface nonlinearities enhanced by plasmon resonances,
in high-index dielectic nanostructures the bulk nonlinearity may dominate the optical nonlinear interaction in volume, giving rise to better conversion efficiency.
In Ref.~\cite{Carletti2015} the disk-shaped nanoantenna made of aluminum gallium arsenide (AlGaAs) is designed for SHG at near-IR wavelengths.
The authors consider this choice of material favorable because AlGaAs possesses a large volume quadratic susceptibility and parasitic TPA effects can be avoided at wavelengths close to 1.55 $\mu$m by engineering the Al molar fraction in the alloy composition.
The second-order bulk nonlinear susceptibility tensor of AlGaAs, possessing a zinc blende crystalline structure, is anisotropic and contains only off-diagonal elements $\chi^{(2)}_{ijk}$ with $i \ne j \ne k$. Thus, in the principal-axis system of the crystal, the $i$th component of the nonlinear source polarization at SH frequency is given by
$P^{(2\omega)}_{i} = \varepsilon_0 \chi^{(2)}_{ijk} E^{(\omega)}_j E^{(\omega)}_k$.
Figure~{\ref{fig:fig4}} shows that the maximum SHG efficiency is achieved for a pump wavelength of 1675 nm, which is close to the magnetic dipole resonance wavelength,  % of 1640 nm.
and predicted to reach values as high as $10^{-3}$ at 1 GW/cm$^2$.
The conversion efficiency strongly depends on the overlap between the induced nonlinear source and modes generated at SH frequency, and its peak appears slightly shifted from the resonance seen in the linear scattering. The SH nonlinear current distribution resembles an electric quadrupole in this case, and pronounced quadrupolar lobes can be recognized in the multilobe SH far-field radiation pattern~\cite{Carletti2015,Carletti2016}. 
In the recent experiment, SH conversion efficiency exceeding $10^{-5}$ has been measured for AlGaAs nanocylinders with an optimized geometry pumped at the wavelength of 1.55 $\mu$m~\cite{Gili2016}. 
% the reported dependence of the experimentally detected SHG on the radius of the nanocylinders with the fixed height 
Ref.~\cite{Carletti2016} further theoretically discusses shaping SH radiation profile of AlGaAs cylindrical nanoantennas through the engineered multipolar interference by manipulating the pump beam (polarization state, incidence angle, spatial structure) and by changing the disk geometry. In particular, whereas no SH emission can be generated by the AlGaAs nanodisk in the normal forward or backward direction under normal excitation because of the structural symmetry and properties of the AlGaAs nonlinear susceptibility, using a pump beam at tilted incidence is beneficial for obtaining SH signal in the normal directions and increasing the detectable SH power that can be measured through a finite numerical aperture microscope objective. 

\section*{Concluding remarks and outlook}

We have discussed multipolar nonlinear effects in resonant metallic, metal-dielectric, and all-dielectric photonic structures. The studies of nonlinear phenomena accompanying the propagation and scattering of light in nanostructured media have been stimulated by a rapid progress in nanofabrication % technology
techniques, such as electron beam lithography and wet chemical growth, and a growing interest in photonic metamaterials with optically-induced magnetic response. The approach based on the multipolar expansion of the electromagnetic fields is closely associated with the Mie scattering theory, and it provides an unified description and useful insights into the physics underlying the nonlinear processes with resonantly driven nanoparticles and nanoparticle clusters. The realization of submicron structures of simple geometries exhibiting intense nonlinear response is eminently promising for the optics-based data manipulation and storage technology, sensing applications, and %recently emerged
topological photonics. 

While the extensive theoretical and experimental studies of nonlinear optical properties of metallic nanoparticles laid the foundation of contemporary nonlinear nanoplasmonics, in many cases appealing to the electric dipole resonances, 
all-dielectric systems have recently been demonstrated to offer stronger nonlinear effects and novel functionalities enabled by pronounced magnetic dipole and higher-order Mie-type resonances. The coexistence of strong electric and magnetic resonances and their interference in high-index dielectric and semiconductor nanostructures bring new physics to simple geometries, opening unique prospects for constructing optical nanoantennas and metasurfaces with reduced dissipative losses and considerable enhancement of the nonlinearly generated fields. 
%
% The attractive property of subwavelength particles is a strong connection between far-field scattering and near-field radiation that allows to control light at the nanoscale. 
% The recent advances suggest the emergence of a new branch of nanophotonics, {\em multipolar nonlinear nanophotonics}, that aims at the manipulation of strong optically-induced electric and magnetic resonances in hybrid and high-refractive dielectric nanoparticles for a design of optical nanoantennas and metasurfaces with unique opportunities for reduced dissipative losses and large resonant enhancement of nonlinear fields.
%
These recent developments suggest intriguing opportunities for a design of nonlinear subwavelength light sources with reconfigurable radiation characteristics and engineering large effective optical nonlinearities at the nanoscale, which could have important implications for novel nonlinear photonic metadevices operating beyond the diffraction limit.

%------------------------------------------------------------------------------%
% \section*{Acknowledgements}\label{sec:Acknowledgements}
\vspace{\baselineskip}
{\bf Acknowledgements.}
The authors thank M.~Kauranen, A.~Zayats, H.~Giessen, O.~J.~F.~Martin, N.~Panoiu, V.~Valev and C.~De~Angelis 
for useful discussions and suggestions. This work was supported by the Australian Research Council and RFBR Grant No. 16-02-00547.

%------------------------------------------------------------------------------%
% \bibliographystyle{apsrev4-1}
% \bibliographystyle{osajnl}
% \bibliography{my_all_diel}

%------------------------------------------------------------------------------%

\end{document}